# Some Tectonic Concepts Relevant to the Study of Rocky Exoplanets

Keith D. Putirka

*Dept. Earth & Env. Sciences*
*California State University*
*Fresno, California, 93740 U.S.A.*

## INTRODUCTION

An Yin of UCLA was invited to write this chapter because he had a near unique approach to understanding tectonic processes. Unfortunately, he passed away on a field trip just a month or two before manuscripts were due. His work was informed by both field studies and theory, and applied to very wide-ranging systems, including the Tibetan Plateau (e.g., Murphy et al. 1997; Kapp et al. 2005), low- angle normal faults of the Basin and Range province (Yin, 1989) and, in the planetary realm, the tectonics of Enceladus (Yin et al. 2016) and Mars (e.g., Yin and Wang 2023). An Yin had planned to write a chapter on planetary tectonics; the loss is massive and, in this moment at least, irreplaceable.

In this necessarily imperfect substitute we'll examine plate tectonics on Earth – its features and forces - and examine some concepts that may allow astronomers to ask useful questions regarding numeric models that putatively predict tectonic activity. But exo-planetologists should be aware that geologists are still attempting to understand: why does Earth operates as it does, and so much differently than its neighbors? Has it always operated this way and have other planets of the inner Solar System ever mimicked Earth's behavior in their past? These problems are unsolved, though some interesting speculative notions have emerged. Studies by Foley et al. et al. (2012) and Weller and Lenardic (2018), for example, attempt to distill the essential planetary properties that may influence if not dictate possible tectonic states, while Yin et al. (2016) propose a model of planetary tectonic surface features that appears remarkably precise. These studies yield some compelling expedients for analyses of planetary objects both within and outside our Solar System.

An additional underlying theme of this chapter is to address the terms "Terrestrial" or "Earth-like". In the astronomical literature, such a description often means only that a planet is covered in rocks; with such usage "Earth-like" is then synonymous with "Mars-like", or "Mercury-like" or "Moon-like"—even "asteroid-like", all of which are entirely unhelpful. Rocky bodies of the inner Solar System operate quite differently from one another—and Earth is different in the extreme. For instance, to say that Earth is unique by virtue of exhibiting plate tectonics is also to say many other things as well: Earth has large continents, and a bi-modal distribution of rock types and topography (hypsometry); volcanic and seismic activity are concentrated at plate boundaries and earthquakes can occur at very great depths (often to as great as 660 km but now recorded to 750 km, i.e., within the lower mantle; Kiser et al. 2021). Abundant surface water might be not just a signal but a qualification to allow some or all these features. To say that a planet is "Earth-like" can and perhaps should mean all these things. In isolation, these features might not guarantee that plate tectonics is operative. In combination, though, they either indicate that plate tectonics is active, or otherwise imply that very Earth-like planets can be created by other means.

# SOME DEFINITIONS AND BASIC CONCEPTS

**Plate Tectonics & Its Basic Parts**

The root of "tectonics" is "tektos", meaning "to weave" or "to build", and its use can refer to earthquakes, mountain building or any deformation of a planet's surface. On Earth, tectonics takes on a special – and thus far unique – form, where the rocky surface is broken up into mobile plates and where earthquakes, volcanoes and surface deformation is concentrated at plate boundaries. Plate motions, or "plate tectonics" also control Earth's heat loss, as well as the global water and carbon cycles that affect climate and habitability (e.g., Foley and Driscoll 2016). Several other kinds of "tectonics" are also recognized, which will also be noted.

In plate tectonic theory, the mobile plates are made of ***lithosphere*** (literally "rocky sphere") that can deform by largely by brittle processes, at least at their shallowest depths (with ductile mechanisms occurring in the deeper, hotter parts). The lithospheric plates float on top of a weak layer called the "***asthenosphere***". The lithosphere can be defined in multiple ways: here, we will mostly refer to lithosphere as the "conductive lid", where heat loss is via conduction, as opposed to convection. But the lithosphere can also be defined by its seismic properties or its elastic strength, or even by the length of time that it has been physically isolated from the convecting mantle, and such definitions can lead to different estimates of its thickness. The lithosphere can be treated as a single unit based on its thermal or mechanical properties, but almost always consists of two parts in terms of composition: a crust, or top layer, that is usually richer in $SiO_2$ and poorer in $MgO$, and an underlying mantle layer that has lower $SiO_2$ and higher $MgO$, among other relative qualities. A key issue is that while the crust is compositionally distinct, the lithosphere could be similar in composition to the underlying convective mantle, though it need not be (Rychert et al. 2020). Ancient lithosphere beneath very old continents may also be quite compositionally distinct, especially with respect to its trace element and isotopic composition (Rudnick et al. 1998). The asthenosphere is a part of the convective mantle, and is weak because it is either slightly hydrated or is very close to or slightly above its solidus temperature (Rychert et al. 2020), but as will be discussed, the overlying lithospheric plates are not carried along in conveyor-belt fashion by convective currents. Convection currents are dominated by thermal upwellings called "***mantle plumes***" and downwelling currents called "***subduction zones***" (Davies and Richards 1992). Mantle plumes emanate from the core-mantle boundary, where they obtain their excess heat; they intersect the surface at random locations relative to plate boundaries. The downwelling limbs, i.e., subduction zones are considered the primary driver for mantle convection (Conrad and Lithgow-Bertelloni 2004) and are defined by a special type of plate boundary: where two plates collide, the denser one is "***subducted***" or thrust beneath the less dense plate, possibly sinking as far as the core-mantle boundary (Fig. 1). Crust and lithosphere are thus destroyed at subduction zones, but new crust is created at spreading ridges, where two plates move away from one another. Convective currents, however, have no relation to spreading ridges, except for the case when a thermal plume accidentally rises beneath a spreading ridge, as at Iceland, and the Galapagos Islands (Fig. 1). Two kinds of plate boundaries have now been described: spreading ridges and subduction zones. Examples of a spreading ridge are the Mid-Atlantic Ridge and the East Pacific Rise. A third type of plate boundary is a ***transform boundary***, where two plates slide past one another; the San Andreas fault is a well-known example, but most transform faults occur in the ocean basins, such as the Romanche Fracture Zone; transform boundaries connect offset segments of spreading ridges.

Subduction zones are of special interest because these are the sites where continental crust – another hallmark feature of Earth - is created. As a subducted slab sinks into the mantle it releases water that was absorbed by the rocks when it was once at the surface. This water is released into the mantle wedge, where it lowers the melting temperature of the mantle, which then leads to the creation of a "***volcanic arc***" that occurs on top of the overriding plate (Grove et al. 2012). Earth's largest and deepest earthquakes also occur at subduction zones; the zone of earthquakes called the Wadati-Benioff zone. The Japanese Islands are an example of an "oceanic arc"—a volcanic arc formed when one oceanic plate (usually older, colder and denser) subducts beneath another oceanic plate (usually younger, hotter and more buoyant). The Cascades are an example of a "continental arc", when an oceanic plate (denser) subducts beneath a continental plate (more buoyant, in part because it contains less dense continental crust). The Tibetan Plateau is also a site of a plate collision. The large plateau represents the collision of two continental plates, India and Eurasia; both plates are buoyant, so there is only minor subduction, and much uplift, and even some volcanic activity, as well as large earthquakes.

    The relative motions of the plates control seismic and volcanic activity and the location of deep valleys and tall mountain ranges. New crust is created at **spreading ridges**: these are mostly oceanic, such as the East Pacific Rise, but spreading ridges can split continents (the Red Sea Rift is an example). Where two plates are pulled apart, the mantle below "passively" upwells to fill the gap (McKenzie 1967; McKenzie and Bickle 1988). We use the term "passive" to distinguish this from the active, thermally-driven upwelling that drives the formation of mantle plumes at the core/mantle boundary (Morgan 1971). The passive mantle upwelling at spreading ridges causes the mantle to partially melt, and form the volcanic rock type called basalt (Langmuir et al. 1992). That basalt layer, once formed, has a constant thickness as it moves away from the ridge, but some of the underlying mantle that was partially melted may now be less dense by virtue of having some its melt extracted (Kay and Kay 1993), and this mantle residue can become a part of the oceanic lithosphere; it is made of a rock called peridotite. The entire package of basalt crust + peridotite mantle comprises the lithosphere. The lithosphere will cool and thicken at its base (effectively increasing the mantle portion) as it moves away from a spreading ridge. What we call the "lithosphere", then, consists of both this newly formed, constant thickness crust, as well as the cooler parts of the underlying mantle that are no longer part of the deeper convective mantle. If the lithosphere thickens and cools sufficiently, and collides with less dense lithosphere, it can be subducted back into the mantle.

    As to the physical properties of the tectonic plates, oceanic crust has a thickness of about 7 km and a density of 2.9 g/cm$^3$, and it overlies a cold, brittle mantle lithosphere that is about 3.35 g/cm$^3$ in density. For the continents, crustal density may be closer to 2.7 g/cm$^3$, and the mantle beneath it may be similar to sub-oceanic lithosphere (3.35 g/cm$^3$) or lower, if it is hydrated (see Kay and Kay 1993). Naturally enough, then, if oceanic and continental lithosphere collide, the denser oceanic lithosphere should sink. If two oceanic plates collide, the older (colder) of the two should subduct. The removal of partial melt from the mantle (to form crust) could also lead to a reduction in the density of the residual mantle, which could also then help isolate such residuum from the convective mantle (Kay and Kay 1993). Interestingly, though, once subduction is initiated, the downward pull of the dense, subducting slab exerts the greatest force acting on tectonic plates. This "slab pull" force is controlled by the sinking of cooler (and so denser) subducted lithosphere, facilitated by the conversion of lower-density basalt (ca. 2.8 g/cm$^3$) to higher-density ***eclogite*** (ca. 3.5-3.8 g/cm$^3$, made of clinopyroxene and garnet). Basalt only transforms to eclogite, though, after it has been subducted to some considerable depth (ca.

40-80 km ; Stern 2002). This implies that thermal buoyancy is critical to initiate subduction. Stern (2004) posits an intervening phase of "hinged subduction", where cool lithosphere dips into the mantle.

In any case, as the lithosphere moves away from an oceanic spreading ridge, it cools, and if carried far enough from the ridge, the base of the lithosphere can be as deep as 100 km or more. Continental lithosphere could be even thicker; some estimates are as high as high as 250-400 km, for very old, cold lithosphere that has been isolated from the convecting mantle for billions of years (Pollack and Chapman 1977; Jordan 1978; Steinberger and Becker 2016), although modeling of surface stresses appears to only allow lithosphere that is mostly 100 km thick on average (Naliboff et al. 2012). The overlying continental crust is just a small portion of this, averaging 30 km in thickness, but ranging to 40-60 km in some places, such as the Tibetan Plateau. Being buoyant, this crust and it's underlying lithosphere can be resistant to subduction.

**The Alternatives: Delamination, Stagnant Lid (or Plume Lid or Drip) & Episodic Lid Tectonics**

A *stagnant lid* "mode" is the case when planets do not exhibit a mobile lid ("mobile lid" being a synonym for plate tectonics). The lithosphere, in this case, acts as a single, globe-encircling plate. With no plate boundaries, earthquakes are few and volcanism and seismic activity are not concentrated in regional bands. Mercury, Moon and Mars are prime examples (Tosi and Padovan 2021). A current stagnant lid regime, though, does not preclude that plate tectonics operated during some earlier era.

A stagnant lid regime also does not preclude surface deformation, or the recycling of some of the lower crust or upper mantle back into a planetary interior. Bird (1979) coined the term "*delamination*" to describe the process whereby the cold mantle beneath the continents may be sufficiently dense to drip back into the mantle. Deep crustal rocks could also delaminate, if they recrystallize at high pressures to form eclogite (Fig. 2a; see also Kay and Kay 1993). Ducea and Saleeby (1996) and Saleeby et al. (2003) demonstrated the reality of such processes beneath the Sierra Nevada mountain range, which led to this new form of tectonic activity, involving gravitational instabilities, to be widely accepted. Building on work by Kay and Kay (1993) and others, Ducea and Saleeby (1996) and Saleeby et al. (2003) also show how delamination can occur as a natural consequence of crustal growth at an arc, as the roots of arc volcanoes can contain very dense mineral assemblages that are gravitationally unstable. Lee and Anderson (2015) coined the term "arclogite" to describe such dense rocks (see also Ducea et al. 2021), and De Celles et al. (2009) argue that the process is repeatable and perhaps cyclic.

More importantly to the study of exoplanets are "lithospheric drips". These might provide a means to initiate subduction (e.g., Adam et al. 2021) on an otherwise stagnant lid planet but also may be the primary means of convection when lids remain stagnant. The crusts on smaller planets might not reach sufficiently high pressures to form eclogite; Batra and Foley (2022) quantify such effects and the consequences for planetary cooling. On a stagnant lid planet, delamination, if it is mechanically possible, may provide the only means by which crustal materials can be recycled back into the mantle. This kind of return flow, and the surface deformation that ensues, has been referred to as "stagnant lid-", "plume lid-" or "drip-" tectonics (e.g., Fischer and Gerya 2016; Stern et al. 2018; McMillan and Schoenbohm 2022), but all involve the same fundamental process of delamination. Yet another variation of activity, perhaps not properly considered "tectonic", is the "heat pipe" model (Moore 2001), an idea that derives from the study of Io, a moon of Jupiter, where volcanism occurs absent large-scale mantle convection. Heat pipes might be restricted to planets with tidal heating and so would be a factor in exoplanets that are in close orbit to their stars (Jackson et al. 2008). An example would be the Trappist-1 system (Barr et al. 2018). In any case the heat pipe model will be ignored further here.

A variation on the stagnant lid regime derives from ideas about tectonism at Venus. To explain the very young surface there (< 500 Ma), Turcotte (1993) applied the term "plate tectonics" but described something different, i.e., that the Venusian crust was subject to catastrophic replacement, as the crust sinks into the mantle and the planet undergoes volcanic re-surfacing. Venus nominally exhibits a stagnant lid between these catastrophic episodes. The resurfacing involves a type of subduction, but something quite different from the stable, arc-like subduction zones on Earth. Modeling of the Venusian processes (Moresi and Solomatov 1998) would lead to the term "episodic lid" (e.g., Nakagawa and Tackely; 2015; Tian et al. 2023) to describe this mode of tectonics.

**Why one and not another?**

We don't know. The idea of a stagnant lid goes back at least as far as the early 1990s, as a result of experiments that simulate mantle convection (Giannandrea and Christensen 1993). In such experiments (though not necessarily in Nature), a stagnant lid regime results when colder surface layers are more than $10^3$-$10^4$ times greater in viscosity compared to underlying, convective materials (Giannandrea and Christensen 1993; Stevenson 2003). With very large viscosity contrasts, a stagnant lid develops on top of a convecting layer; with lower viscosity contrasts, or in effect, a weaker lithosphere, the surface layers become a part of the convection process, and so a mobile lid regime ensues. This all seems simple enough, except that as shown by Solomatov (1995), and later noted by Stevenson (2003), Earth is an anomaly relative to such experimental results: Earth has a very large viscosity contrast between its lithosphere and asthenosphere (ca. $10^9$; Doglioni et al. 2011), and yet still exhibits plate tectonics.

Moresi and Solomatov (1998) propose that it's not so much the viscosity of the lithosphere that matters as much as the frictional resistance to brittle failure that dictates when and where plate tectonics happens. We've long known, for example, that water can greatly reduce friction along fault surfaces (Hubbert and Rubey 1959), making such faults much weaker than their drier counterparts. The Hubbert and Rubey (1959) model would explain, at least qualitatively, why plate tectonics occurs on Earth, where water in the crust and lithosphere is abundant, but not on Venus, where the Venusian crust and lithosphere are dry. However, as noted by Bercovici et al. (2015) and emphasized by Foley (2018), the predicted lithosphere yield stresses in current numerical models (e.g., Moresi and Solomatov 1998;Tackely et al. 2000) require yield stresses that are orders of magnitude lower than observed (e.g., Zhong and Watts 2013) so as to allow plate tectonics, at least if the unidentified units employed by such models are in Pascals. It remains a challenge to bring observations and predictions of stress into agreement.

Yet another challenge is for numerical models to predict the narrow width of deformation zones. Wakabayashi (2021), for example, shows that in the Franciscan Formation of California (an archetype of an exhumed subduction zone), field and microstructural evidence indicate a zone of deformation that is a mere 300 m wide, and that this narrow deformation interval applies over a depth range of 10 to 80 km in the paleo-subduction zone. However, even very recent models appear challenged to obtain a plate boundary that is realistically narrow. Saxena et al. (2023), for example, appear to recognize that deformation zones are narrow, as they prescribe "discrete weak zones" within their numerical model, but their zones of weakness are still hundreds of km in width, orders of magnitude greater than the actual deformation widths that occur on Earth.

The effective rheological contrasts that control stagnant/mobile lid behavior also depend on gravity, and whether partial melt is present in the convecting layer, and there might also be a stochastic dependency—noted by Stevenson (2003) as "the vagaries of history" or by Lenardic and Crowley (2012) as "historical dependence". The vagaries are not entirely vague, though. Rock strength can depend upon strain history, as anyone who has tried straightening a bent metal wire can attest (when rocks are strained, or wires bent, microscopic dislocations are created to accommodate the strain, but these dislocations can later interfere with one another adding to intrinsic strength). Volcanic activity can also dehydrate a planet's interior (drier rocks are stronger) and rock strength can be affected by recrystallization and crystal size (Evans and Kohlstedt 1993). Lenardic and Crowley (2012) thus illustrate how different models, using very similar inputs, can yield different results regarding whether or not a planet exhibits plate tectonics.

In yet another approach to explaining plate tectonics, Hoink et al. (2012) suggest that the stagnant vs. mobile lid regimes are controlled not so much by the properties of the lithosphere and underlying bulk mantle, but instead depend upon the thickness of a low viscosity "channel" that directly underlies the lithosphere. If this channel is sufficiently thin then, presumably, plate tectonics, or a mobile lid regime is sustained, whereas thick channels promote a stagnant lid regime. However, this model predicts strong viscous coupling for some plates, such as at the Mid-Atlantic Ridge, despite both long-standing and recent geophysical observations to the contrary: plate motions show that slab pull accounts for 90% of the forces that drive plate tectonics, with ridge push accounting for most of the remainder, and with viscous forces being trivial (Chapple and Tullis 1977; Hager and O'Connell 1981; Lithgow-Bertelloni and Richards 1998; Conrad and Lithgow-Bertelloni 2004; Saxena et al. 2022). Conrad and Lithgow-Bertelloni (2004) do indeed describe a viscous suction force that can act on the deeper parts of subducted plates, but their viscous suction is another form of slab pull and is not an argument for viscous coupling at shallower depths. Clennett et al. (2023) propose that viscous forces might have been more important in Earth's past, but also concede that uncertainties in plate reconstructions are large enough to disallow that possibility. That lithospheric plates might be riding upon mantle convection cells is an alluring idea—but there is no evidence that the idea applies to Earth.

Finally, while mantle convection plays a key role in modeling of plate tectonics, some questions remain about the operational aspects. Mantle plumes are the upwelling complement to subduction, and are thought to be thermally driven, with their excess heat derived from the core-mantle boundary (Davies and Richards (1992). But Frazer and Korenaga (2022) suggest that Earth's mantle plumes are much too wide (by about 400%) compared to what is predicted in numerical models. They conclude that either thermal convection models are wrong or that convection is driven more by compositional, rather than thermal contrasts. The latter result of compositionally-driven convection, though, would upend our views of how plumes nucleate and how Earth's core loses heat.

In sum, we have a mismatch between observations of plates and plumes on the one hand and numerical models that attempt to predict these on the other. And so we find ourselves at a loss to precisely explain plate tectonics and mantle convection, on a planet where observations and relevant experiments are abundant.

# PLATE TECTONIC DRIVING FORCES

Several developments in the late 20th century helped to solidify our understanding of the mechanical aspects of plate tectonics. One of the most important was by McKenzie (1967) who showed that heat flow and bathymetry measurements at mid-ocean spreading ridges require no source of excess heat in the underlying mantle. Mid-ocean ridges are a product of "passive upwelling" of the mantle, where two plates are being pulled apart due to far-field forces and the mantle passively upwells to fill the gap (Fig. 1). Mid-ocean ridges (MORs) are bathymetrically high relative to surrounding ocean floor because cold, dense lithosphere is replaced by warm and buoyant mantle. That buoyant mantle provides heat sufficient to explain heat flow measurements. But MORs are not so high so as to require or allow excessively hot, buoyant mantle material beneath them (McKenzie 1967). Passive upwelling means that at a given depth (ca.100 km) below the surface, temperatures are more or less uniform under the ocean basins, being no higher beneath a spreading ridge than laterally adjacent regions. The mantle is by no means homogeneous with respect to density and temperature (Langmuir et al. 1992; Adam et al. 2021), but the idea of passive upwelling fits well with the cooling model for Earth's oceanic lithosphere. For the case of conductive cooling, the depth to the base of an oceanic plate, $d$, can be described as a square root relationship with the age of the plate, $t$:

$$d = \sqrt{\kappa t} \qquad (1)$$

where $\kappa$ is thermal diffusivity; $t$ is determined by age dating of oceanic rocks, but can be also calculated as $t = L/v$, where $L$ is the length of a plate, and $v$ is plate velocity. The bathymetry of the ocean basins, at least to $t = 90$ Ma, is proportional to the square root of the age of the overlying oceanic crust, as predicted by equation 1 (Parsons and Sclater 1977). At ages >90 Ma, the ocean floor bathymetry is shallower than predicted by equation 1, and can be explained by small-scale convection in a narrow boundary layer beneath the oceanic lithosphere. This long-standing view has been tested repeatedly and is consistent with inferred temperature profiles of the lithosphere and observations of gravity (Morgan and Smith 1992; Crosby et al. 2006).

These observations, and modeling of such, show that tectonic plates are not driven conveyor-like on mantle convection currents, but rather are controlled by other forces. Forsyth and Uyeda (1975) presented an analysis of plate-driving forces that are still accepted today, namely ridge push, viscous drag and slab pull, with the slab pull force being the most important and viscous drag the most minor. Empirical observations (e.g., Jarrard 1986) and at least some numerical models that are specifically designed to predict modern plate motions (e.g., Conrad and Lithgow-Bertelloni 2002, 2004; Saxena et al. 2023) confirm that analysis, showing that the slab-pull force, which is generated by a subducted plate's negative buoyancy, is as great as 60%-70% of the total forces acting on a given plate. The fraction of the slab pull force increases with decreasing mantle viscosity, and the remainder is still a type of slab pull, but referred to as "slab-suction" (Conrad and Lithgow-Bertelloni 2002, 2004). The suction force is derived from mantle flow around the sinking plate, where that flow can exert some shear along the plate that exerts an additional downward force. The suction force is nominally more effective in the deeper mantle where mantle viscosity is greater. Ridge push comprises <10% of the total force acting on plates and viscous coupling is effectively zero. These particular models are consonant with seismic observations: Ito et al. (2014) show that there is only a weak coupling of the lithosphere with underlying mantle plume beneath Hawaii, and the Jadamec (2016) analysis of mantle anisotropy shows rather complex mantle flow that is de-coupled from subducted slabs, at least to depths of 100 km.

Davies and Richards (1992) describe the issue well, noting that if mantle upwelling were the driving force at mid-ocean ridges, then all submarine ridges would look more like Iceland, with much steeper and shallower bathymetry than observed, as well as higher heat flow (McKenzie 1967) and a positive geoid anomaly (Sleep 1990; the "geoid" is the shape the Earth would have if it were covered in water, or any other fluid that lacked shear strength; above mantle plumes, as at Hawaii or Iceland, there is a positive geoid anomaly in that Earth's shape is distended upwards above the rising jet of mantle material). The idea of plates being moved about on the top of convective mantle rolls is limited to dated textbooks and some computer models; it is not a feature of the modern Earth.

Convection in the mantle, then, appears to be dominated by large plumes that upwell from the base of the mantle, receiving their excess temperatures from the core-mantle boundary (CMB), and sinking convective limbs that are defined by the subducting plates, much as illustrated in Silver et al. (1988) and Davies and Richards (1992). These plumes have no spatial correlation to the mid-ocean ridges; their surface manifestations are the so-called volcanic hot spots, such as Hawaii, or Yellowstone, where the excess heat drives partial melting and volcanic activity. Mantle plumes might, by accident, intersect a mid-ocean ridge, as appears to happen at Iceland or Galapagos, and can create thicker crust in such areas. But these are accidents that do not bear directly upon plate motion. The mantle plume theory is an essential addition to plate tectonic theory in that it explains volcanic and seismic activity that occurs far from plate boundaries, and so is otherwise anomalous in plate tectonic theory. But mantle plumes and plate tectonics are not unrelated, at least on Earth.

**Buoyancy Forces, Rock Types, & Isostacy**

Mantle circulation and plate tectonics are largely driven by buoyancy forces. Rocks can differ in composition, temperature or both, affecting their bulk density. Plate motions and mantle circulation can ensue when the resulting buoyancy forces are sufficient to overcome ambient rock strength. The Rayleigh number, Ra, is one measure to assess whether a planetary interior might convect. There are many different variations; this equation is from Stevenson (2003):

$$Ra = \frac{g\alpha\Delta T d^3}{\nu\kappa} \qquad (2)$$

where $g$ is the acceleration due to gravity, $\alpha$ is the coefficient of thermal expansion, $\Delta T$ is the temperature drop across some depth interval $d$, $\nu$ is the kinematic viscosity and $\kappa$ is thermal diffusivity. When $Ra > 1000$ the buoyancy forces in the numerator sufficiently exceed the resistive forces in the denominator and the system is likely to convect. As a planet cools, $\Delta T$ will decrease, and when convection stops ($Ra < 1000$), heat loss would then be dominated by conduction.

Stevenson (2003) interestingly shows, however, that smaller planets can apparently cool just as slowly as their larger counterparts—maintaining a lower $T$ through their cooling history (Stevenson, 2003). Basalt compositions from the Moon and Mars appear to confirm this view (Putirka 2016). In any case, Stevenson (2003) is largely concerned with convection as a mode of heat flow, noting that planets may instead lose heat via conduction and melt migration if very small (Moon and possibly Mercury), or via heat pipes and magma oceans if very hot (early Earth), and will lose heat by convection otherwise.

We are uncertain of the extent to which density contrasts in Earth's mantle are driven by compositional or thermal contrasts, as seismic velocities within Earth's interior can be explained by either. Adam et al. (2021) indicate that known thermal contrasts of ca 300°C (as derived from thermometry of volcanic rocks at the surface) may explain most density contrasts within Earth, which would then allow for a well-mixed mantle that might be homogenous at certain spatial scales. But compositional contrasts clearly dominate near Earth's surface and appear to drive plate tectonics. There are many different rock types that drive such differences, but two, granite and basalt, are of particular importance as the former occurs only in great abundances on continents while the latter dominates the ocean basins. As noted by Campbell and Taylor (1983), only Earth contains multiple, large "batholithic masses" of granite (a recent discovery indicates a possible batholithic mass on Moon; Siegler et al. 2023). Some definitions are needed here. Campbell and Taylor (1983) refer to "granite", *sensu lato*, i.e., any rock that is dominated by the minerals quartz (Qz; $SiO_2$), albite (Ab; $NaAlSi_3O_8$), orthoclase (Or; $KAlSi_3O_8$) and anorthite (An; $CaAl_2Si_2O_8$). Granitic rocks cool slowly enough that the crystals tend to be large and easily identifiable. The latter three minerals, Ab, Or and An, are "feldspars" and form two different solid solution series: the alkali feldspars (Afs), which are a mixture of Ab-Or (Na and K are close in size and have a similar charge so mix relatively easily) and the plagioclase feldspars (Plag), which are a mixture of Ab-An ($Na^{1+}$ and $Ca^{2+}$ are similar enough in size but differ in their charge, so they can mix together via a coupled substitution: $Na^{1+}Si^{4+} = Ca^{2+}Al^{3+}$). The larger $K^+$ and smaller $Ca^{2+}$ ions are sufficiently different in size so as to preclude mixing between these two series. Other minerals are often present in granite, especially hornblende and biotite, but Qz, Afs and Plag dominate (>90%). "Granite", *sensu stricto*, represents very specific limits of the relative abundances of these minerals; when they are normalized to 100%, Qz ranges between 40-60 %, Afs is generally greater than 10%, and the remainder is Plag (Les Bas and Streckeisen 1991). In this chapter, we will consider the case of "granite" *sensu lato*, unless specified otherwise. Basalt, on the other hand, is a volcanic rock the cools quickly – sometimes so quickly that much of the rock consists of glass. But usually basalt is not entirely glass, and the minerals olivine, clinopyroxene and An-rich feldspar are very common. Chemically, granite is rich in $SiO_2$ and low in MgO, while basalt has the opposite characteristics. The differences in composition and mineralogy lead to differences in density: granite and basalt have ranges in density of ca. 2.6-2.7 and 2.8-3.0 $g/cm^3$ respectively.

As abundant as Earth's continental crust is relative to our planetary neighbors, it still represents just 1/3 of Earth's surface. The remaining 2/3 of Earth's crust lies beneath the oceans and is made of basalt, which is created by partial melting of Earth's peridotite mantle (Langmuir et al. 1992). But while a 2/3 coverage area may seem impressive, basalt dominates to an even greater extent the surfaces of other planets, comprising effectively 100% of the crusts of Moon, Mars and Mercury, and the parent bodies of all non-chondritic meteorites. It probably also covers much of Venus as well, though the surface of Venus is still largely unknown.

In contrast to basalt, granite has a more complex origin that appears to involve water – and plate tectonics (Grove et al. 2012). Campbell and Taylor (1983) note that some of our neighboring planets might have small occurrences of granite, but none have granite in "batholithic masses" i.e., the occurrences are not so large so as to define a large mountain range let alone a continent. Again, some definitions: an individual blob of granite, perhaps a few km in diameter or smaller, is called a ***pluton***; in many places, such as the Sierra Nevada mountain range of California, many thousands of plutons of similar age are emplaced, one against the next, to form a near continuous belt of granitic rocks that dominate a mountain range of hundreds of km in extent; these collected masses of granite form a **batholith**. The tectonic significance is that granitic batholiths appear to form beneath continental volcanic arcs (Fig. 1), and so are intimately related to subduction. The association between subduction and granitic batholiths is related to the global water cycle (see Grove et al. 2012): (a) near a mid-ocean spreading ridge, water is able to penetrate the basaltic crust and into the peridotite mantle below, creating a wide range of hydrous minerals; (b) as the hydrated lithosphere (basalt crust + underlying peridotite lithosphere) is subducted, many of the hydrous minerals break down into anhydrous minerals, releasing water to the mantle wedge (Fig. 1), which also consists of peridotite; (c) the water released to the mantle wedge decreases the melting point of peridotite and partial melting ensues; (d) the melts so produced are water-rich and poor in Si, and as these melts rise upwards, they precipitate amphibole, among other phases; (e) experiments (Sisson et al. 2005) show that the precipitation of amphibole ($SiO_2$-poor) yields residual liquids (magmas) that are enriched in $SiO_2$, producing granite. Campbell and Taylor (1983) hypothesize that occurrences of surface water, batholiths and plate tectonics are so intimately linked that all these might have initiated on Earth together or in close temporal association. (Foley and Driscoll (2016) take such connections further, showing that tectonic activity on Earth has other cascading effects, in yielding a global C cycle and extended cooling of the core so as to respectively influence Earth's climate and perpetuate a core dynamo).

Compositional (density) contrasts also explain Earth's rather unique topography: Earth's continents sit at higher elevations than the ocean basins, giving Earth a distinctly bi-modal topographic distribution. These elevation differences can be modeled via the concept of isostacy (Fig. 2b), where the oceanic and continental crust (and its underlying rocky mantle lithosphere) floats upon a plastically deformable part of the upper (asthenosphere). The asthenosphere has a lower viscosity than the lithosphere, which means that any lateral differences in pressure within the asthenosphere can be alleviated by lateral flow. The pressure at any depth within Earth is close to $P = \rho g h$, where $P$ is pressure, $g$ is the acceleration due to gravity and $h$ depth. If the mean $\rho_1$ for a column or rock differs from the mean $\rho_2$ for an adjacent column of rock, the relative heights, $h_1$ and $h_2$ of the two columns, can adjust until the pressure at the "depth of compensation" (sometimes, but not necessarily within the asthenosphere) is equalized, so that $P_1 = P_2$. Columns with lower mean $\rho$ will rise to higher elevations compared to those with higher average $\rho$—provided that the effective buoyancy forces exceed ambient rock strength—and in the process yields what is known as isostatic equilibrium.

**How the Mantle Flows and Rocks Break**

To model plate tectonics, we study mantle circulation and lithospheric strength. The bulk properties (yield strength, viscosity) are obtained from experiments on rocks and minerals. The experimental conditions could include a fixed differential stress and strain rate. The differential stress, $\sigma$, is determined from the difference between the maximum (usually noted as $\sigma_1$) and minimum (usually noted as $\sigma_3$) principal stresses (so $\sigma = \sigma_3 - \sigma_1$) that are applied in a given experiment.

To understand mantle convection, i.e., plastic deformation within the interior of a planet, we can use equations that involve strain rate ($\dot{\varepsilon}$), which measures how rapidly a rock is deformed. For example, if a rock has an initial length of $l_o$ and a final length of $l_f$, then its total strain is $\varepsilon = (l_f - l_o)/ l_o$. If that strain occurs over a time, $t$, then the strain rate is $\dot{\varepsilon} = \varepsilon /t$. Various "constitutive laws" describe how strain rate is related to differential stress, grain size, and other variables. For example, if deformation in a rock is controlled by diffusion (so the dominant means of plastic deformation involves the migration of "vacancies", which are un-occupied atomic sites), then a constitutive law, from Evans and Kohlstedt (1993), is:

$$\dot{\varepsilon} = A \frac{\sigma^n V}{RT} \frac{D}{d^m} \quad (3)$$

Where R is the gas constant and T is temperature; $\sigma$ is the differential stress as just defined, and its exponent, *n*, is nominally close to 1; *d* is the average grain size and its exponent, *m*, is nominally equal to 3, if diffusion is fastest along grain boundaries, but is equal to 2 if diffusion is fastest through the grain itself; *D* is the diffusion coefficient, and both *D* and the pre-exponential coefficient *A* can depend upon temperature (*T*), pressure (*P*) and water content, which in some forms of the equation are explicitly shown. In Evans and Kohlstedt (1993) the values for *A*, *n* and *m* for diffusion creep respectively vary from $10^{-3}$ MPa$^{-n}$/s, 1.1 and 3.0 for dunite (a rock made solely of the mineral olivine) to $10^{4.9}$ MPa$^{-n}$/s, 1.7 and 1.7 for calcite. Earth's mantle consists of ca. 60% olivine ($[Mg,Fe]SiO_4$), and calcite ($CaCO_3$) is thought to be absent, so mantle deformation should be better described by the first set of values. Another mode of deformation is called dislocation creep, where plastic deformation is accommodated by the migration of dislocations (in effect, micro-faults at the atomic scale); a general constitutive law for dislocation (or "power law") creep, adapted from Evans and Kohlstedt (1993), can be written as:

$$\dot{\varepsilon} = A\sigma^n exp\left(\frac{-Q}{RT}\right) \quad (4)$$

where R is the gas constant, T is temperature and Q is an activation energy; for dislocation creep, the exponent *n* on the $\sigma$ term is generally > 1, and its value may denote a specific style of dislocation creep.

Evans and Kohlstedt (1993) tabulate experimental values of *A, n* and *Q* for rocks that are dominated by particular minerals (see Table 1). The mantle is dominated by olivine but also contains up to 20% or more each of orthopyroxene (Opx; $[Mg,Fe]_2Si_2O_6$) and clinopyroxene (Cpx; $Ca[Mg,Fe]Si_2O_6$). Some mantle rocks approach nearly 100% pyroxene, and pyroxenes have different strengths compared to olivine (Table 1). These differences in strength can depend on temperature. For example, Yamamoto et al. (2008) and Bystricky et al. (2016) both find that at low *T*, Opx is stronger than Ol, and Bystricky and Mackwell (2001) find that Cpx is stronger than olivine, so long as water is absent. A fascinating study by Hansen and Warren (2015) appears to at least qualitatively substantiate these experimental findings in a natural setting, where pyroxene-rich rock samples deformed at ca. 1000°C are 1.2 to 3.3 times more viscous than olivine-rich samples. However, Bystricky et al. (2016) note that the contrasts in the strengths of olivine and pyroxenes may decrease with increased *T*. If valid, this would imply that shifts in mantle mineralogy are less important for understanding mantle convection than for lithosphere (tectonic plate) strength.

**Table 1.** Power Law Creep constants for various minerals and rock types

| Mineral | *A (MPa$^{-n}$/s)* | *n* | *Q (kJ/mole)* | *Source[1]* |
|---|---|---|---|---|
| Olivine | $10^{-1} - 10^{5.4}$ | 2.1-5.1 | 226-540 | E&K93 |
| Clinopyroxene | $10^{10.8}$ | 4.7 ±0.2 | 760 | B&M |
| Orthopyroxene | $10^{8.63}$ | 2.8-3.1 | 583-621 | B2016 |
| Quartz - dry | $10^{-11.2} - 10^{-2.9}$ | 1.9 - 11 | 51 - 377 | E&K16 |
| Quartz - wet | $10^{-9.4} - 10^{-1.4}$ | 1.4 - 4 | 134 - 230 | E&K93 |
| Calcite | $10^{-3.6} - 10^{8}$ | 3.3-8.3 | 190-427 | E&K93 |

[1]E&K93 = Evans and Kohlstedt (1993), B2016 = Bystricky et al. (2016), and B&M = Bystricky and Mackwell (2001).

The very wide-ranging values reported by Evans and Kohlstedt (1993) are not just a product of experimental uncertainty: they reflect varying experimental conditions, which include contrasts in crystal growth rates and final grain size, porosity, pressure, temperature, the presence or absence of water or melt (if the experiments are conducted at high $T$), and the maximum value of $\sigma_1$. Recent experiments (Hansen et al. 2019), for example, show that the strength of olivine increases with decreasing grain size and that there is also significant strain hardening (as dislocations develop to a sufficient degree so as to interfere with one another, thus requiring greater stresses to induce deformation). If such potential variations are not, in and of themselves, sufficient to give one pause, imagine the havoc that ensues if any exoplanet might trap significant amounts of $CO_2$ so as to stabilize calcite in their otherwise olivine+pyroxene-dominated mantle.

Consider also that olivine is not stable below depths of about 400 km, as it transitions into the minerals β-phase spinel (having the same composition as olivine but different structure), wadsleyite and ringwoodite. Olivine could be critical to understanding the brittle behavior of tectonic plates themselves, but we are only just now beginning to understand the physical properties of the remaining 2500 km of Earth's 2980 km-deep mantle. New experiments by Fei et al. (2023), for example, indicate that the lower-mantle mineral, bridgmanite, grows at a faster rate, and so produces larger crystals than other minerals that occur at mid-mantle depths (ca. 800-1,200 km). Fei et al. (2023) also find that, despite other grain-sized trends to the contrary, this particular trend to larger crystals can explain a $10^{1}$-$10^{1.3}$ factor of viscosity increase below 800 km, close to inferred values (Rudolph et al. 2015).

**Brittle Deformation and The Plate Boundary**

The deformation described by Eqns. (3-4) can be distributed over thousands of km, but in sharp contrast, deformation with the crust, and perhaps even within parts of the deep lithosphere, can occur in very narrow bands. We have already noted the findings of Wakabayashi (2021), which shows that deformation zones within subduction zones is concentrated in a zone of < 300 m. This narrowness has also been documented elsewhere. The San Andreas Fault is one of the largest and best-known transform faults, and for much of its 1300 km length, fault offsets occur on sub-parallel faults that may have a total width of no more than a few hundreds of meters. Drill cores into the fault zone (the San Andreas Fault Observatory at Depth or SAFOD) show that deformation within this particular fault strand is concentrated within bands of <200 m wide, and is frequently as narrow as 2-3 m (Zoback et al. 2011). These results are remarkably similar to the very narrow deformation widths determined within fossil subduction zones. Strain hardening was noted as a possible effect in dislocation creep, but in fault zones, the narrowness of these zones may be an effect of strain weakening (e.g., Gueydan et al. 2014).

In any case, the equations that describe fault behavior are naturally different than those that describe plastic deformation during mantle convection. In the colder, brittle lithosphere, fault motion is called "stick-slip", which describes the type of discontinuous forces that are needed to overcome frictional resistance along a fault surface. Figure 3a shows a case where the maximum principal stress ($\sigma_1$) is vertical, while the intermediate ($\sigma_2$) and minimum ($\sigma_3$) principal stresses are horizontal. If the differential stress ($\sigma = \sigma_1 - \sigma_3$) is sufficiently great a failure plane can develop, as indicated by the shaded surface (Fig. 3a). Figure 3b shows the stresses in 2-dimensions, where the two principal stresses can be resolved into a normal stress, $\sigma_n$, oriented perpendicular to the failure plane, and a shear stress, $\tau$, oriented parallel to the plane of failure. In a solid rock, with no pre-existing fractures, failure occurs when $\tau$ is greater than the sum of the shear strength of the rock ($S_o$), in the absence of a confining pressure and with a correction for the increased resistance to failure as confining pressure increases; this last pressure-sensitive term is written as the product of the normal stress on the fault plane and an "internal coefficient of friction", $\mu$, so the rock strength is:

$$\tau = S_o + \mu\sigma_n \qquad (5)$$

When there is a pre-existing plane of weakness, then failure occurs when the shear stress is greater than the product of the normal stress and coefficient of friction along the failure plane, $\mu_s$:

$$\tau = \mu_s\sigma_n \qquad (6)$$

An analysis of experimental data (Byerlee 1978) shows that rock strength can be approximated as:

$$\tau = 0.85\sigma_n \qquad P < 2 \text{ kbar} \qquad (7a)$$
$$\tau = 0.5 + 0.6\sigma_n \qquad 2 \text{ kbar} < P < 20 \text{ kbar} \qquad (7b)$$

which together are known as "Byerlee's Law".

## SO WHEN IS A PLANET EARTH-LIKE?

From a geologic point of view, an Earth-like planet must have plate tectonics, where the tectonic plates (lithosphere) experience brittle deformation, concentrated within remarkably narrow bands. Beneath the lithosphere would be a convecting mantle that is dominated by buoyant thermal plumes and cold lithospheric plates that are subducted back into the mantle. Such a planet would have a bi-model hypsometry with high elevation continental crust and low-lying oceanic basins. The continental crust would owe much of its buoyancy to a significant fraction of low-density rocks that have andesitic and granitic compositions, with quartz and feldspar-dominated mineralogies. The ocean basins would consistent of denser, basaltic rocks, with olivine and feldspar minerlaogies. The continental crust would be a product of magmatism that occurs above subduction zones (arcs) while the oceanic plates would result from volcanism at rifts in the lithosphere. Rifting would have no relation to convection within the deeper mantle but would instead be a passive response to the slab pull forces that drive subduction. Actively upwelling parts of the mantle – thermal mantle plumes - would intersect the oceanic spreading ridges only by coincidence and would drive their own form of volcanism at the surface. Earth-like plate tectonics likely requires water at a planetary surface and thus also implies a global water cycle, as water that is subducted can be returned to the surface during arc-related volcanism.

# PROBLEMS OF SUBDUCTION

Subduction and the related formation of granitic crust at arcs are hallmarks of an Earth-like planet, and probably key to identifying an Earth-like planet elsewhere in the galaxy. Only Venus appears to have subduction (Schubert and Sandwell 1995; Devaille et al. 2017); it appears to be a minor feature, though Turcotte (1995) has proposed that Venus might lose heat via a catastrophic and episodic replacement of its crust. Tian et al. (2023) refers to Venus as an "episodic lid" regime. But this type of wholesale subduction, interposed by long periods of quiescence, would still yield a surface that can be near uniform in age. A varied and tectonically young surface would appear to be a terrestrial (as opposed to a Venusian) hallmark. If Earth-like subduction, with well-defined and abundant arcuate zones of volcanism and seismic activity, could be identified on any rocky planet outside of Earth, it would be a spectacular discovery and perhaps a signal of Earth-like tectonics that might not be possible via an episodic lid regime. But the question of how and why any particular long-lived subduction zone is initiated is an open question (see Stern 2004). No less crucial is the question of how and when plate tectonics is initiated.

Estimates for the initiation of terrestrial plate tectonics vary from >4 to < 1 Ga. Parman et al. (2001) suggested that certain volcanic rocks from the Archean (4.0-2.5 Ga) are derived from subduction, and their study would thus place subduction initiation on Earth at least as far back as 3.5 Ga—a mere 1 Ga post accretion. Stern (2005) would later argue that the rock types and assemblages that characterize modern subduction zones do not appear in the rock record until the Neoproterozoic (1 – 0.54 Ga), which would make subduction a relatively recent evolutionary event. Yin et al. (2020), though, can push the rock record of modern subduction back to at least 2 Ga. These claims are not necessarily mutually exclusive: it's possible that plate tectonics was operative in the Archean, but created different rock types on a hotter Earth. Brown et al. (2019) suggest something along these lines, noting that that the absence of the rock types that Stern (2002) notes could be less telling than the existence of bi-modal temperature-pressure ($T/P$) regimes: subduction zones yield rocks with relatively cold temperatures at high pressures, whereas a "normal geotherm" yields rocks that record high temperatures at high pressures. Brown et al. (2019) suggest that bi-modal $T/P$ conditions can be identified in metamorphic rocks at least as early as the Neoarchean (2.8-2.5 Ga). Putirka (2016) suggested a similar test, noting that arc-related volcanic are produced at lower temperatures, and so volcanic activity that is thermally bi-modal should indicate plate tectonics, whereas thermally unimodal volcanism would indicate a stagnant lid; Earth's volcanic rock record indicates bi-modality in the Archean.

An especially intriguing set of observations derive from study of the oldest mineral grains yet discovered on Earth. Harrison et al. (2005) examine zircon grains ($ZrSiO_4$) from Australia that have been aged dated at 4.0-4.4 Ga. These ancient zircons yield wide ranging isotopic ratios, and the parent and daughter isotopes are known to fractionate through crust-forming processes that occur at subduction zones. The nature and magnitude of their isotopic shifts indicate that the amounts of continental crust in the Hadean Era (4.5-4.0 Ga) could have been close to modern values, and that some residues of the crust-forming events were subducted. Other compositional aspects of these zircons indicate that they formed in a wet, low-temperature environment, not unlike modern granitic rocks that form at arcs. The data are not perfectly conclusive (Harrison 2009), but important nonetheless. A key counter-argument derives from Bernard (2006), who suggests that granitic rocks on the Archean Earth could form within a plume-lid (stagnant lid) tectonic regime, where thermal plumes yield a thickened crust that can then delaminate repeatedly to generate new granitic rocks—and a very thick continental crust, absent subduction. However, a new study by Ge et al. (2023) makes a compelling case for Archean plate tectonics as they show that Archean crustal zircons record greater amounts of $H_2O$ and higher $fO_2$ compared to their coeval mantle counterparts; such rocks are less oxidized than modern analogs, but the crust/mantle contrasts are the same, and are easily explained if water and O are being subducted into the mantle to create the Archean crust.

It is not clear that numerical models can yet play a role in informing the issue of the start of plate tectonics on Earth. Earth was assuredly hotter in the first 1-2 Ga after accretion, and it unclear whether such a state of affairs has the effect of causing the viscosity contrast to increase (by decreasing mantle viscosity) or decrease (by decreasing the lithosphere viscosity via heating from the mantle). Numerical experiments by Foley (2018) indicate no dependency of plate tectonics on interior mantle temperature since convection can yield weak plate boundaries, by a form of strain weakening, at a range of temperatures. Foley (2018) interestingly finds that plate motions are more sluggish in a hotter Earth, which may be unsurprising if cold, dense slabs drive plate motion.

What does this all have to do with exoplanets? Putirka and Xu (2021) argued that the discovery of granitic rocks on an exoplanet might provide the surest evidence of Earth-like behavior. The Bedard (2007) model casts doubt on this idea. The discovery of continental crust might only tell us that we have a tectonically active planet (still interesting), but it might not tell us what type of tectonics is operable. The best tests of exoplanet tectonic styles will derive only from better understanding of the early Earth, or perhaps from the identification, via space telescopes, of water in a planetary atmosphere, that together with granite, could indicate a global water cycle.

## PLATE TECTONICS ELSEWHERE IN THE SOLAR SYSTEM?

It is generally accepted that, besides Earth, plate tectonics does not operate on any of the inner planets (e.g., Devaille et al. Tosi and Padovan 2021). Katterhorn and Prockter (2014) suggest that the icy shell of Europa might have spreading ridges and subduction. All other objects with an observable solid surface in our Solar System would be in a stagnant lid regime (e.g., Stern et al. 2018). Smrekar et al. (2023) describe Venus as a "squishy lid"—but this is still just a subtle variation on a stagnant lid regime. The tectonic behavior of other planets, though, is inferred from surface deformation and their interpretations are non-unique. Besides being stagnant lid planets, Moon and Mercury have something else in common: their surfaces show planet-wide contractional features, which appear to record cooling and contraction of the planetary interior (e.g., Matsuyama et al. 2021; Watters et al. 2016). (This interpretation is, ironically, also the now-discarded explanation for mountain ranges on Earth). Some contractional features on Mercury might also result from mantle down-wellings (Watters et al. 2015), while on Mars, contractional deformation could also involve subduction-like activity, with lava flows piling up against one another (e.g., Yin and Wang 2023).

There has been speculation that Mars had more Earth-like tectonics very early in its history. Yin (2012) showed that subduction may have been active at a local scale to create extension in the Tharsis region. And Connerney et al. (1999) discovered magnetic lineations that appear analogous to those found at oceanic spreading ridges, while Connerney et al. (2005) appear to have discovered transform faults that might offset nominal spreading ridges. Since plate tectonics is expected to speed up heat loss, and perhaps facilitate thermal convection of the core (which nominally creates a planet's magnetic field), Nimmo and Stevenson (2000) suggested that the magnetic fields recorded in the Martian crust might also record the timing of when plate tectonics was active on this planet. But Breuer and Spohn (2003) show that the Martian crust and its magnetic field can be better explained if Mars was always in a stagnant lid regime. And new experiments by Zhou et al. (2022) indicate that the Martian crust might never have been sufficiently dense, relative to its underlying mantle, to subduct. Thermal contraction thus appears to be the cause of most tectonic features on Mars (Nahm and Schultz 2011).

Venus may be the more intriguing case study of tectonics and should provide a testing ground for numerical models that nominally predict tectonic behavior. Venus has a young surface (compared to Moon, Mars and Mercury) and, as noted earlier, the planet is possibly subject to catastrophic resurfacing (Turcotte 1995). Venus also shows distinct patterns of surface deformation that are reminiscent of subduction (DeVaille et al. 2017). Numerical models, such as those by Armann and Tackley (2012), compare a stagnant lid to an episodic lid regime and in preferring the latter, they make very specific predictions about the geology of Venus. For example, compared to the stagnant lid case, an episodic lid regime predicts the number of catastrophic re-surfacing events (between five and eight) and the length of time for each event (ca. 150 Ma), as a thin crust and low heat flow from the core. Tian et al. (2023) present another testable prediction, i.e., that local volcanic resurfacing intervals depend upon crustal rheology. Two questions immediately arise: (a) might any other numerical models yield similar predictions? (b) do the model predictions survive further scrutiny (i.e., geologic mapping) of the Venusian surface?

Numerical models have provided some insights; for example, they emphasize that to convert a stagnant lid into a mobile one, considerable zones of weaknesses must be introduced into the lithosphere. But we kind of knew this anyway, by observation. In any case, any model that is simultaneously capable of predicting the surface features and nominal tectonic behaviors of Earth, Venus, Europa, Moon and Mercury should stand a very good chance of providing useful insights regarding tectonic behavior on exoplanetary bodies—or even the early tectonic history of the early Earth. It is unclear that any such numerical models but progress is possible if we aim towards such challenging goals.

# PROSPECT FOR MODELING TECTONICS ON EXOPLANETS

As with mineralogy (Putirka, this volume), studies of exoplanetary tectonic regimes is largely a game of "what if"; predictions of tectonic behavior in even highly simplified cases are perhaps even less clear than mineralogy estimates, and is perhaps best described as 'informed speculation'. The most profitable means to understanding tectonic regimes is mostly like that of Foley et al. (2012) and Weller and Lenardic (2018), who map the mechanical properties of planets so as to separate plate tectonic and stagnant lid regimes. Their approach has the advantage of producing testable hypotheses and consistency tests. The paper by Foley et al. (2012) is nominally about "super Earths" but is more interesting. As in earlier studies, Foley et al. (2012) argue that planets can exhibit plate tectonics when the viscosity ratio between the lithosphere and mantle, $\mu_l/\mu_m$, is sufficiently low, but in their model, Stevenson's (2003) anomaly is avoided in that they account for strain-weakening in the lithosphere. They do so by considering a critical viscosity ratio $\eta_{lcrit}$:

$$\eta_{lcrit} = \left(w(bRa^{3\beta})^2\right)^{\frac{m}{p}} \tag{8}$$

where *Ra* is the Rayleigh number (similar to Eqn. 1), and *b, m* and *p* are material constants, analogous to the *m* and *n* values in Eqns. 3-4 (in their preferred model, *m* = 2, *p* = 3, *b* = 0.05 and *β* = 1/3); *w* is the more critical parameter here in that it quantifies strain weakening. Plate tectonics is allowed when $\mu_l/\mu_m < \eta_{lcrit}$ (Fig. 4a). If, for whatever reason, *w* is sufficiently high (e.g., water infiltrates Earth's lithosphere), then so might be $\eta_{lcrit}$, which can then offset a high bulk $\mu_l/\mu_m$. Weller and Lenardic (2018) attempt a similar kind of map, using only the yield strength of the lithosphere, but they also consider the percentage of heat generated by radioactivity (*Q*) at any given time in a planet's history (Fig. 4b).

It is unclear to what extent predictions from the two models might differ. But as noted by Foley (2018) there nonetheless remain fundamental disagreements as to the physical models that predict how plate boundaries form. Figure 4b shows that hotter planets are more likely to exist in a stagnant lid regime. The parameter *Q* in Fig. 4b seems, at least superficially, to preserve the idea of a viscosity ratio effect: higher *Q* at a constant lithosphere yield strength implies a higher ratio of $\mu_l/\mu_m$, since higher mantle temperatures would mean a lower $\mu_m$, and higher *Q* expands the stagnant lid regime. Some entrepreneur need only collect mechanical estimates of the various known objects of our Solar System and compare results. But the mechanical properties of even Earth and Venus appear to be in doubt. Weller and Lenardic (2018) plot Earth and Venus at a similar (ca. 50 MPa) yield strength, though the drier Venusian lithosphere seems as if it should be placed somewhat higher (Fig. 4b). However, the authors note that the positions of the two planets are "illustrative only". Especially interesting is that Foley et al. (2012) and Weller and Lenardic (2018) both have Earth teetering at the boundary of the mobile and stagnant lid regimes, despite evidence that Earth has unreservedly exhibited a mobile lid for at least half (Yin et al. 2020), if not all its history.

Note, however, that the mobile/stagnant lid boundary in the model of Foley et al. (2012) is not sensitive to *T* (Fig. 4a). This contrast in results is a critical underlying issue. As Foley (2018) shows, the *T*-sensitivity of the boundary in Fig. 4b can stem from use of a "pseudoplastic rheology" (e.g., Tackley et al. 2000; Korenaga 2010); in such a model plate boundaries form when stresses in the lithosphere exceed a value adopted for the lithosphere's yield strength; at high *T*, stresses are lower, and so may decrease below the yield strength threshold. The problem is that such models require yield strengths orders of magnitude below those observed in nature or experiment, even at low temperature conditions where stresses are higher. Bercovici and Ricard (2012) alternatively propose that plate boundaries form due to a particular style of grain diminution; their model builds on related ideas of strain localization, where a decrease in grain size yields a feedback loop that yields concentrated zones of weakness (e.g., Bercovici and Karato 2003). In the model of Foley (2018), grain diminution has the effect of increasing diffusion creep, which is highly sensitive to grain size (Eqn. 3). The result is that internal mantle temperatures do not control tectonic activity (Foley 2018). This has the salutary effect of allowing what seems likely: plate tectonic activity in the Archean (Ge et al. 2023).

A remaining challenge is to describe surface deformation from thermal states and mechanical properties. Yin et al. (2016) provide an intriguingly precise model of the Saturnian moon Enceladus. That moon has a stagnant lid that exhibits an array of strike-flip faults; the Yin et al. (2016) model, building on studies of terrestrial strike slip faults on Earth (Roy and Royden 2000), predicts the brittle and elastic thicknesses and the cohesive strength of the icy Enceladean lithosphere.

Is there some way to combine predictions of tectonic states and surface deformations? We of course have no immediate prospect of surveying exoplanet surfaces, but models that can predict many things are better than those that predict few. Or is a predictive model a chimera? Foley (2018) shows how there is no clear consensus on the physics of how plate boundaries form. Weller and Lenardic (2018) suggest that, due to feedback loops and model non-linearities, tectonic states and planetary evolutionary paths may be unpredictable. Did Earth evolve to its present state as an accident of its initial conditions, or by stringing together a series of unlikely events? Perhaps. Of Hazen et al.'s (2015) choice between Chance and Necessity, mineral assemblages of planetary mantles and crusts appear to fall to the latter, as the laws of thermodynamics appear determinative. Maybe deformation and tectonics is stochastic, subject to the "vagaries of history". But there are many experiments in rock mechanics still to be performed and much to be explored of the physics of deformation before admitting defeat.

Yin A, Wang Y-C (2023) Formation and modification of wrinkle ridges in the central Tharsis region of Mars as constrained by detailed geomorphological mapping and landsystem analysis. Earth Planet Phys 7, 161-192

Yin A, Zuza AV, Papplardo RT (2016) Mechanics of evenly spaced strike-slip faults and its implications for the formation of tiger-stripe fractures on Saturn's moon Enceladus. Icarus 266, 204-216

Zhong S, Watts AB (2013) Lithospheric deformation induced by loading of the Hawaiian Islands and its implications for mantle rheology. J Geophys Res 118, 6025-6048

Zhou WY, Olson PL, Shearer CK, Agee CB, Townsend JP, Hao M, Hou MQ, Zhang JS (2022) High pressure-temperature phase equilibrium studies on Martian basalts: implications for the failure of planet tectonics on Mars. Earth Planet Sci Lett 594, DOI 10.1016/j.epsl.2022.117751

Zoback M, Hickman S, Ellsworth W, SAFOD Science Team (2011) Scientific Drilling Into the San Andreas Fault Zone – An Overview of SAFOD's First Five Years. Sci Drill 11, 14–28, https://doi.org/10.2204/iodp.sd.11.02.2011## Figure Captions

**Figure 1.** Illustration of plate tectonic features and mantle convection on Earth. Mid-ocean spreading ridges and the transform faults that offset such are shown, as is a subduction zone, the dewatering of the subducted slab that drives partial melting of the mantle wedge, and the volcanic arc that is produced as a result. Also shown is a hot upwelling plume that is rooted at the core/mantle boundary (the only significant thermal boundary layer below Earth's surface) that also drives partial melting in the mantle and volcanism at the surface (at so-called "hot spots"). Note that convection is dominated by plume upwellings and downwelling tectonic plates; this whole-mantle flow has no connection to spreading ridges. Spreading ridges are a result of passive upwelling of the mantle as the plates are ripped apart by far-field forces, but are not connected to the deeper, hotter mantle. See Silver et al. (1988) and Davies and Richards (1992).

**Figure 2**. (a) Delamination and subduction: at the base of either the continental or oceanic crust, the rock type eclogite ($\rho = 3.6$ g/cm$^3$) may form. Being more dense than the mantle below, an eclogitic root beneath a continent may delaminated into the mantle (a form of "drip tectonics"). At the base of oceanic crust, which is already more dense than continental crust, the eclogite drip may initiate subduction, where the entire plate is pulled into the mantle (Fig. 1). (b) Example of isostacy. Low density continental crust ($\rho = 2.8$ g/cm$^3$) will have a "root" that extends to greater depths than adjacent, oceanic crust ($\rho = 2.9$ g/cm$^3$) of higher density. The depth of that root is the depth at which lithostatic pressures (like hydrostatic, pressures are equal in all directions) are equal, which is called the "depth of compensation".

**Figure 3**. Illustration of brittle (Coulomb) failure. (a) Principal stresses $\sigma_1$, $\sigma_2$ and $\sigma_3$ act upon the shaded surface where the rock will eventually break (or perhaps already has broken). Failure is determined by the magnitude of the differential stress, $\sigma_1 - \sigma_3$ and for un-broken rock, by the intrinsic strength of the rock, often referred to as the coefficient of internal friction. (b) The principal stresses can be resolved into a normal stress, $\sigma_1$, and a tangential stress, $\tau$. For a pre-existing fault surface, failure occurs when $\tau$ exceeds the frictional resistance. That frictional resistance can be greatly reduced when water is present, or (at least typically) when the materials near the fault surface recrystallize to smaller grain size.

**Figure 4**. Therman and mechanical criteria that allow plate tectonics. (a) From Foley et al. (2012), critical viscosity, $\eta_{crit}$, is normalized to the viscosity ratio, $\mu_l/\mu_m$, where $\mu_l$ is the viscosity of the lithosphere and $\mu_m$ is the viscosity of the convective mantle. When $\mu_l/\mu_m$ is high (greater than a critical value as determined by $\eta_{crit}$, or $\mu_l/\mu_m > \eta_{crit}$) a stagnant lid regime ensues. The ratio is expected to vary with temperature, as indicated by the solid curve. (b) In the map developed by Weller and Lenardic (2018) only the lithosphere strength is used for the vertical axis, but this is compared to surface temperature and the percentage of heat that is still being produced by radioactivity ($Q$; as opposed to residual heat from accretion and core formation) as a planet ages.

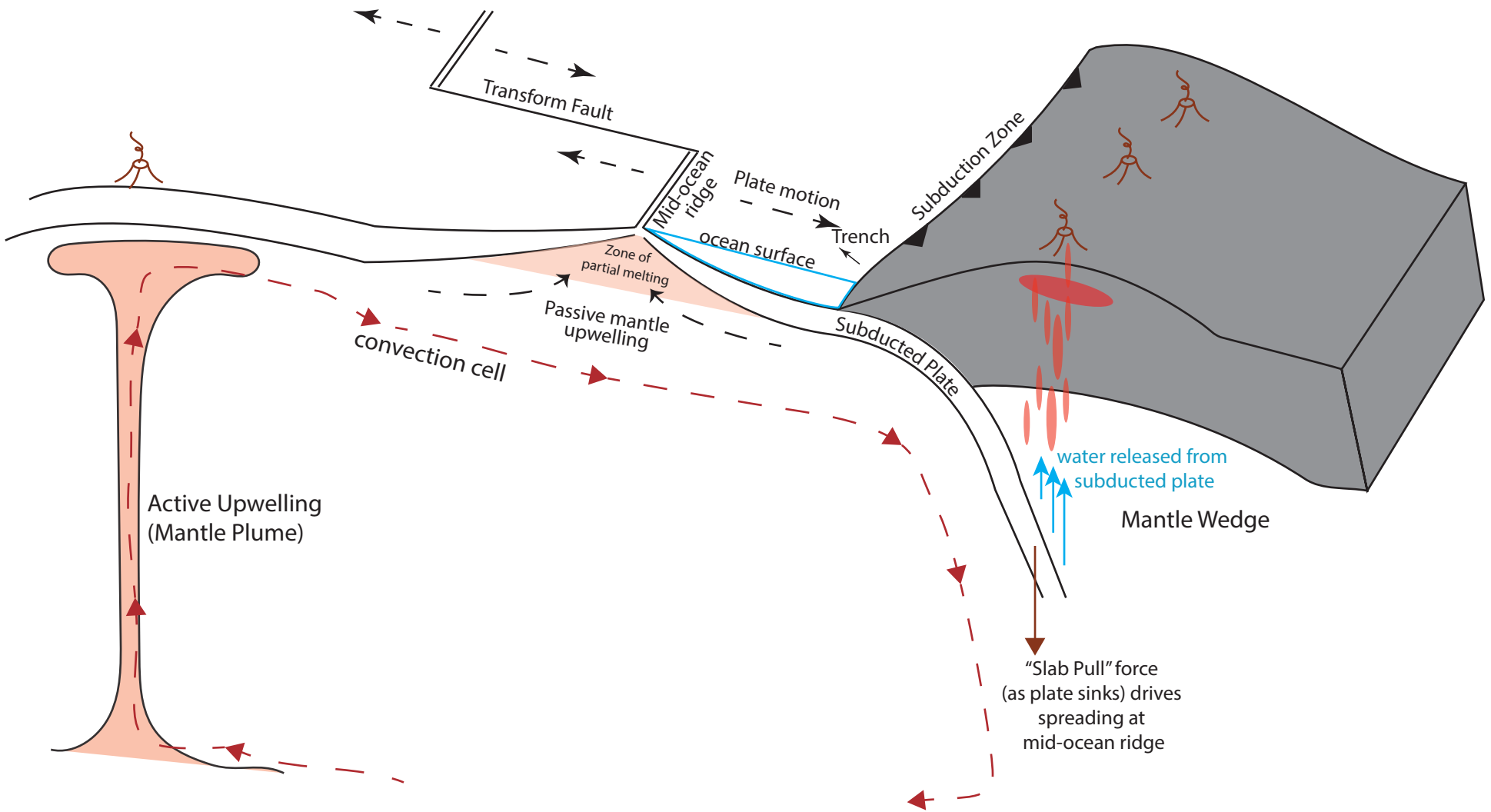

Figure 1

Figure 2

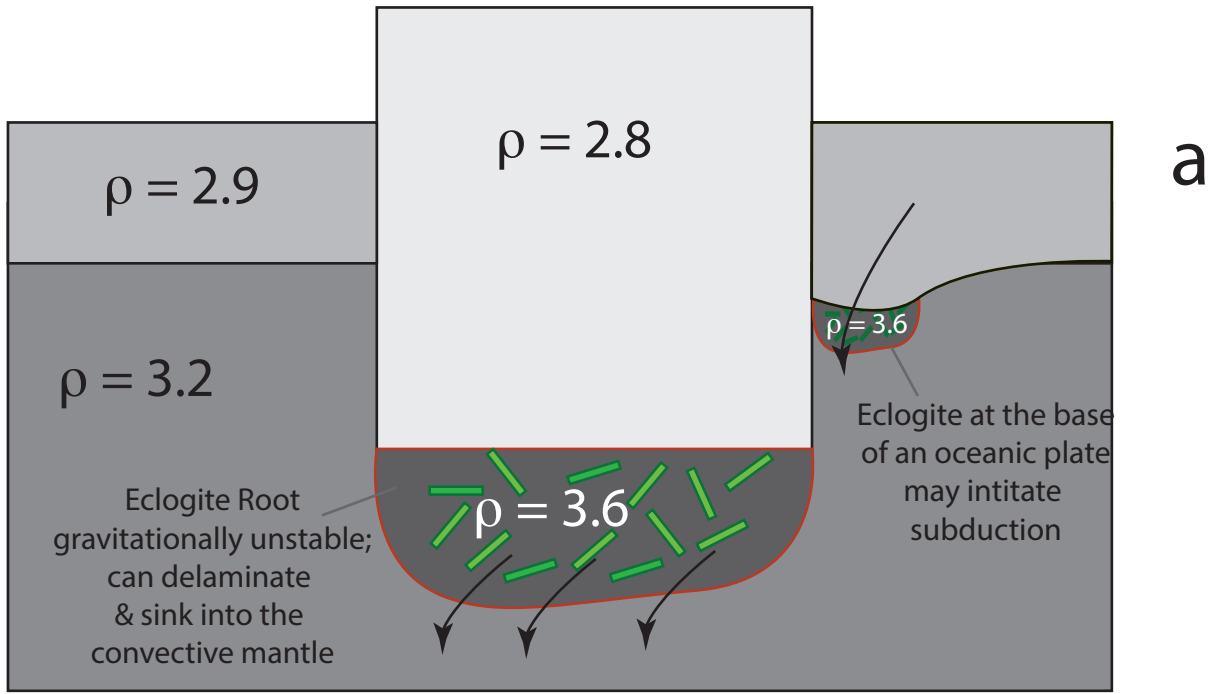

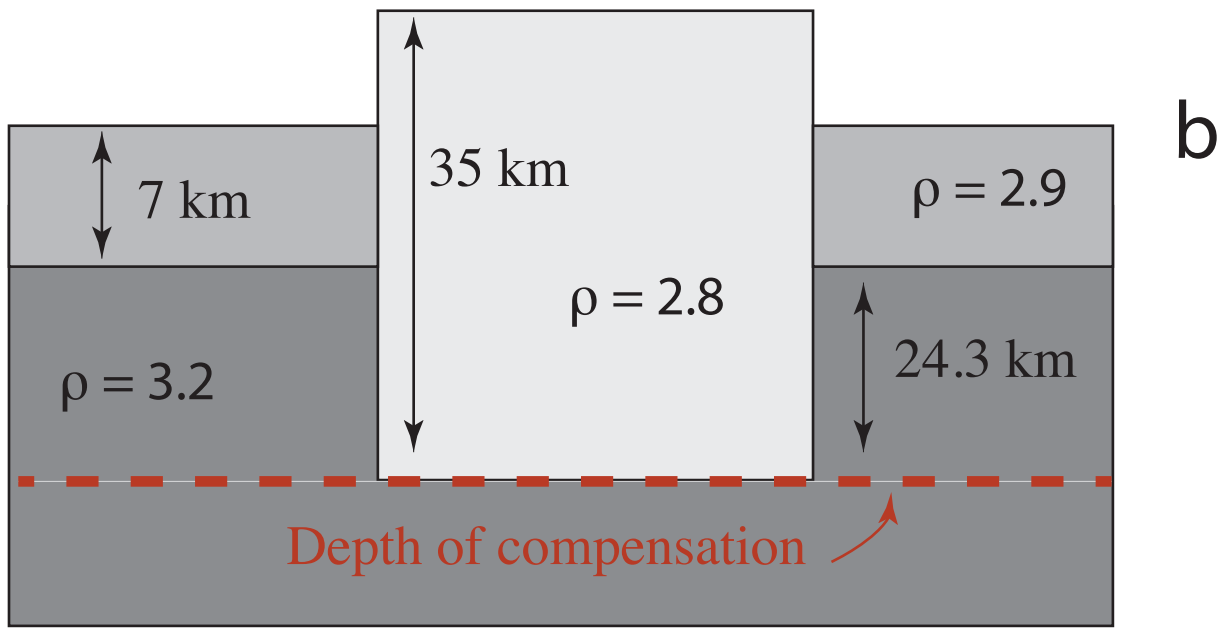

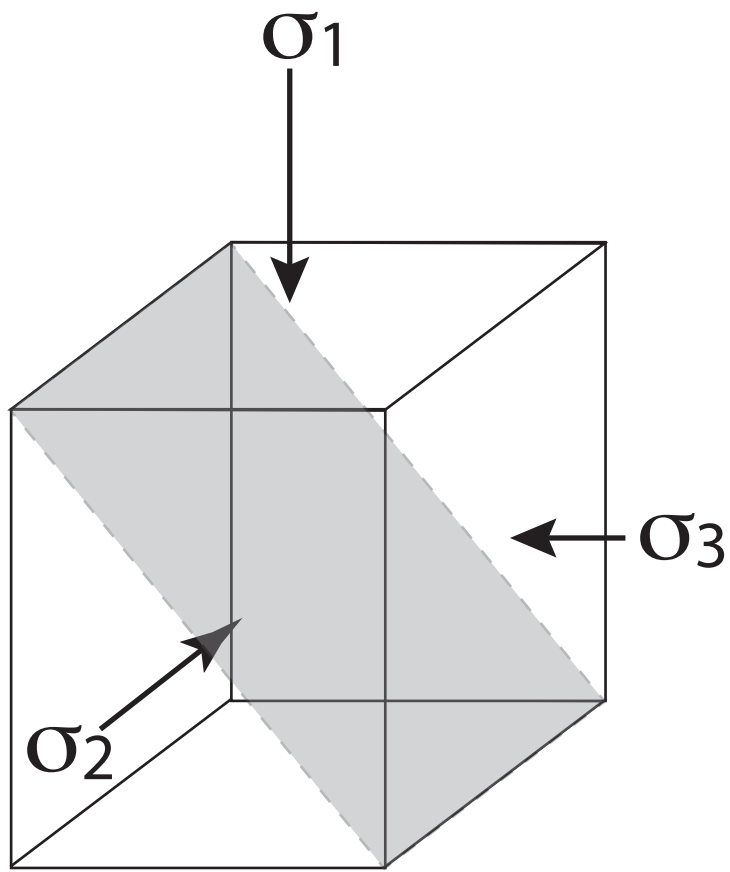

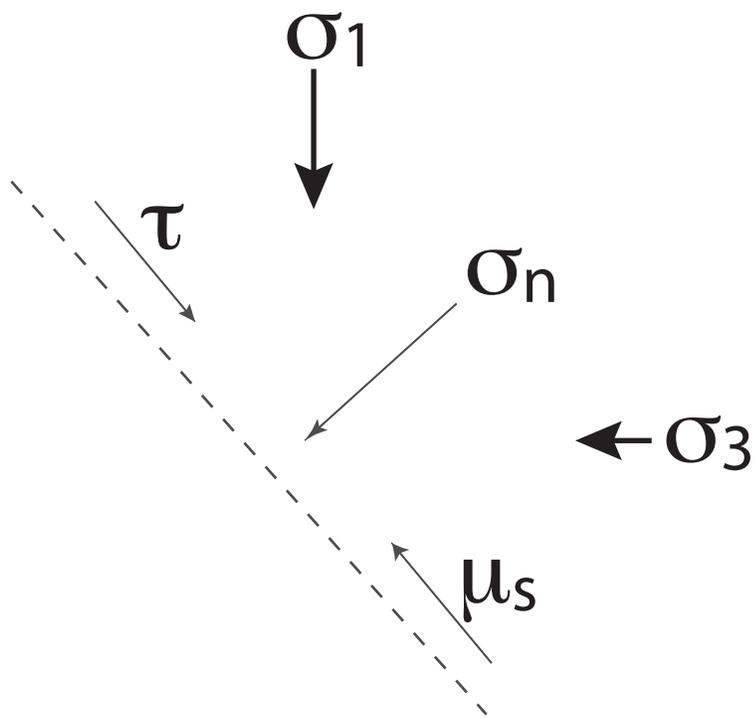

Figure 3

Figure 4

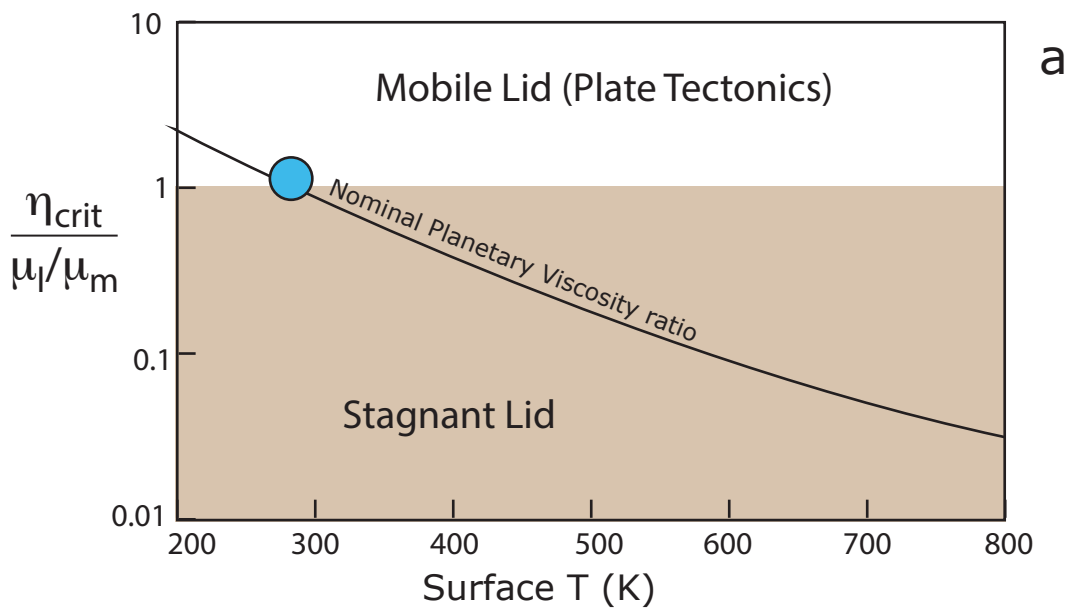

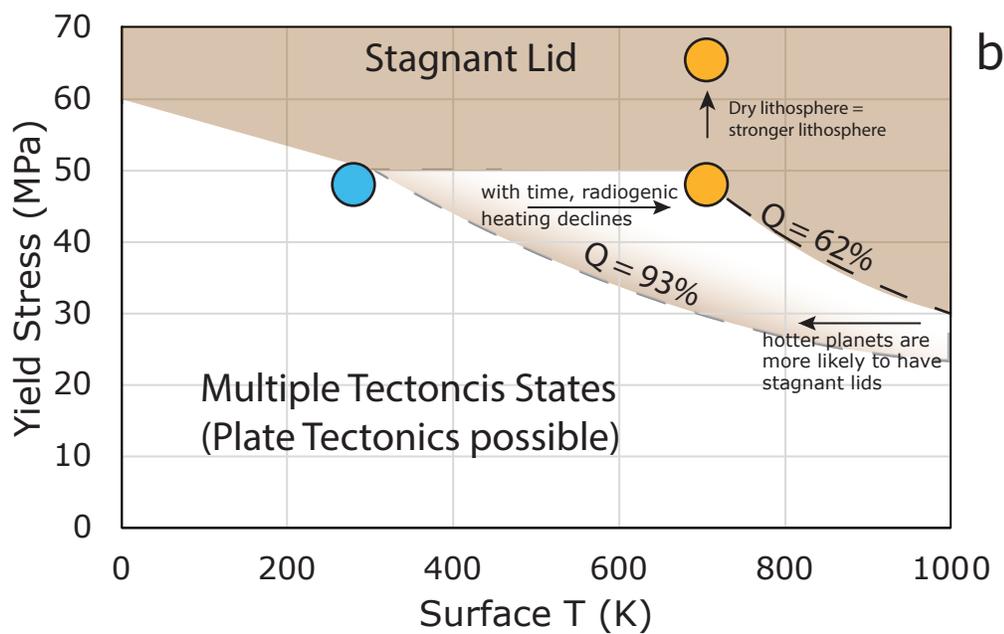